\documentclass[journal=nalefd,manuscript=article]{achemso}

\usepackage[version=3]{mhchem} 
\usepackage[T1]{fontenc}       
\usepackage{graphicx}
\usepackage{units}
\usepackage{dcolumn}
\usepackage{amsmath,amssymb}
\usepackage{verbatim} 
\usepackage[usenames,dvipsnames]{color}
\usepackage{bm} 

\author{Luca Banszerus}
\affiliation[]
{JARA-FIT and 2nd Institute of Physics, RWTH Aachen University, 52074 Aachen, Germany}

\author{Michael Schmitz}
\affiliation[]
{JARA-FIT and 2nd Institute of Physics, RWTH Aachen University, 52074 Aachen, Germany}

\author{Stephan Engels}
\alsoaffiliation[]
{JARA-FIT and 2nd Institute of Physics, RWTH Aachen University, 52074 Aachen, Germany}
\affiliation[]
{Peter Gr\"unberg Institute (PGI-9), Forschungszentrum J\"ulich, 52425 J\"ulich, Germany}

\author{Matthias Goldsche}
\alsoaffiliation[]
{JARA-FIT and 2nd Institute of Physics, RWTH Aachen University, 52074 Aachen, Germany}
\affiliation[]
{Peter Gr\"unberg Institute (PGI-9), Forschungszentrum J\"ulich, 52425 J\"ulich, Germany}

\author{Kenji Watanabe}
\affiliation[]
{National Institute for Materials Science, 1-1 Namiki, Tsukuba, 305-0044, Japan }

\author{Takashi Taniguchi}
\affiliation[]
{National Institute for Materials Science, 1-1 Namiki, Tsukuba, 305-0044, Japan }

\author{Bernd Beschoten}
\affiliation[]
{JARA-FIT and 2nd Institute of Physics, RWTH Aachen University, 52074 Aachen, Germany}

\author{Christoph Stampfer}
\affiliation[]
{JARA-FIT and 2nd Institute of Physics, RWTH Aachen University, 52074 Aachen, Germany}
\alsoaffiliation[]
{Peter Gr\"unberg Institute (PGI-9), Forschungszentrum J\"ulich, 52425 J\"ulich, Germany}
\email{stampfer@physik.rwth-aachen.de}

\title
  {Ballistic transport exceeding 28~$\mu$m in CVD grown graphene}

\abbreviations{CVD, RIE, EBL}
\keywords{graphene, ballistic transport, CVD, cyclotron radius, mean free path}

\begin{document}





\begin{abstract}
We report on ballistic transport over more than 28~$\mu$m in graphene grown by chemical vapor deposition (CVD) that is fully encapsulated in hexagonal boron nitride. The structures are fabricated by an advanced dry van-der-Waals transfer method and exhibit carrier mobilities of up to three million cm$^2$/(Vs). The ballistic nature of charge transport is probed by measuring the bend resistance in cross- and square-shaped devices. Temperature dependent measurements furthermore prove that ballistic transport is maintained exceeding $1~\mu$m up to 200~K.
\end{abstract}

\newpage
The extraordinarily high charge carrier mobility in graphene, \cite{Bol08} up to 150,000 cm$^2$/(Vs) at room temperature~\cite{Wan13}, has recently triggered the interest in ballistic transport experiments that rely on the principles of electron optics~\cite{Mia07,Tay13,Che07,Sin15,Ric13}. Novel device concepts such as Veselago lenses~\cite{Che07}, Klein-tunneling transistors~\cite{wil15}, ballistic graphene nanoribbons\cite{Bar14} and ballistic rectifiers~\cite{Sin15} make use of the chiral nature of graphene electrons and of their long mean free path, exceeding 10~$\mu$m \cite{Wan13}. Apart from the long mean free path, the scalability of the material is another important aspect when heading for practical graphene-based electronics, including Dirac fermion optic devices. Here, a great challenge is to combine large device structures with high charge carrier mobilities. A very promising and scalable fabrication technique of graphene is chemical vapor deposition (CVD), which has recently made tremendous advances in growth quality~\cite{Li09, Bae10, Li11, Che13} and delamination methods~\cite{Suk11,Gan11,Pet12, Ban15}. Both have led to a substantial improvement of the charge carrier mobilities of large area graphene~\cite{Ban15}. In the present work, we push forward these advancements by demonstrating ballistic transport in CVD-grown graphene that is fully encapsulated in hexagonal boron nitride (hBN). We observe ballistic transport up to 200~K in an one micron-sized cross-shaped structure, and show ballistic transport over 28~$\mu$m in a square-shaped device at 1.8~K. The mean free path reported here exceeds previous values for CVD graphene by almost two orders of magnitude.\cite{Cal14}

\begin{figure*}[htb!]\centering
\includegraphics[draft=false,keepaspectratio=true,clip,%
                   width=0.98 \linewidth]%
                   {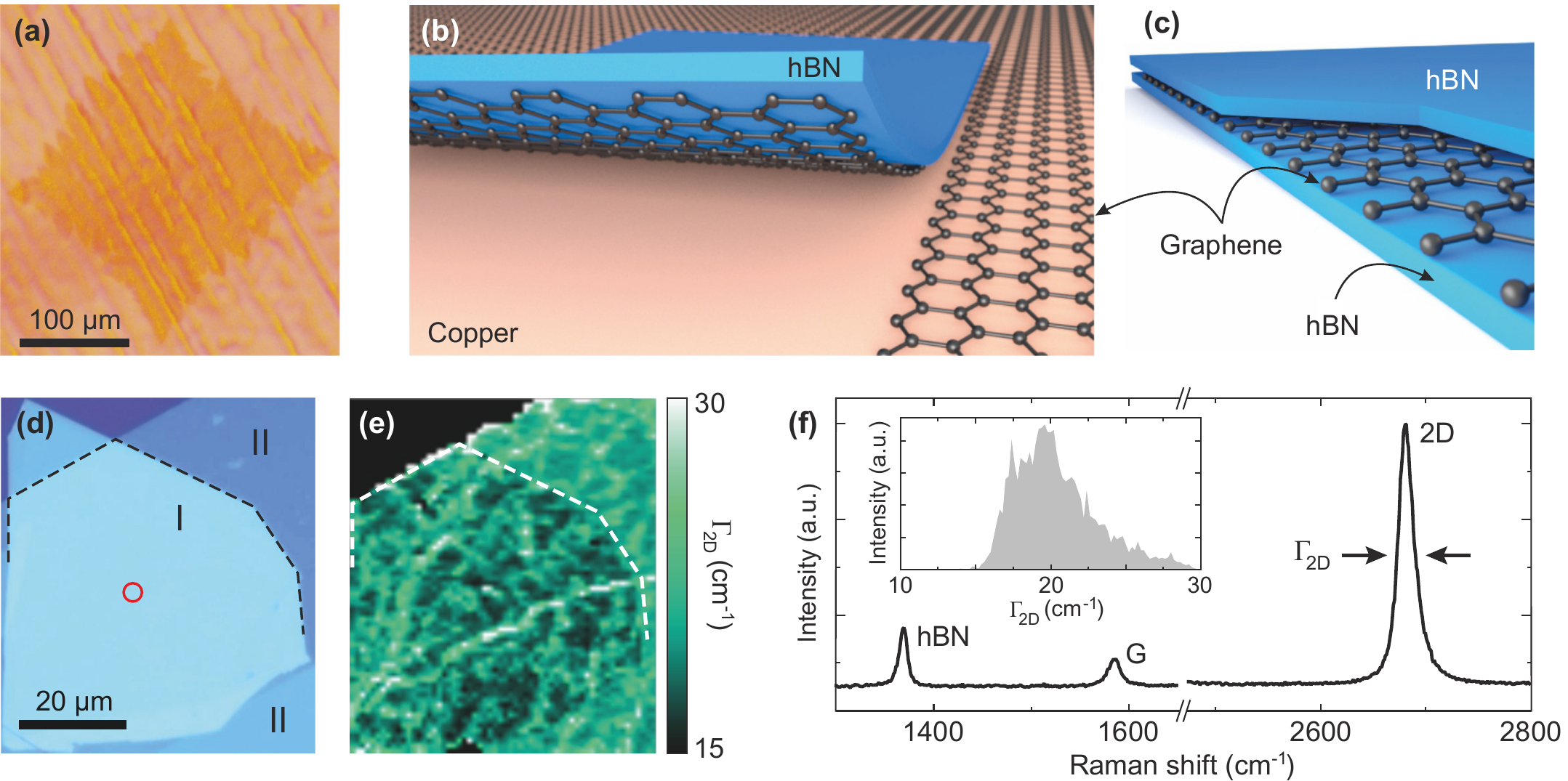}
\caption[fig01]{\textbf{(a)} Optical image of a CVD graphene flake on the catalytic copper foil. 
\textbf{(b)} Illustration of graphene delamination from the copper foil by the van der Waals pick-up by hBN. The graphene sheet is stamped out along the edges of the hBN flake \textbf{(c)}
The hBN/graphene is placed on another hBN flake, resulting in the illustrated sandwich structure.
\textbf{(d)} Optical image of a finished hBN/graphene/hBN heterostructure. In region I, graphene is fully encapsulated by hBN while in region II only the top surface of graphene is protected by hBN.
\textbf{(e)} Raman map of the FWHM of the 2D peak, $\Gamma_{\mathrm{2D}}$, for the sample shown in panel (d). \textbf{(f)} Raman spectrum on fully encapsulated graphene, taken at the position of the red circle in panel (d). The inset shows the histogram of the values of $\Gamma_{\mathrm{2D}}$ recorded over the entire region I.}
\label{fig01}
\end{figure*}

We grow graphene by a low pressure chemical vapor deposition process (LPCVD) within a copper enclosure~\cite{Che13,Ban15}. This process yields large individual single-layer graphene flakes on the copper foil (Figure~1a), with lateral sizes of typically few hundred micrometers. The graphene flakes are subsequently encapsulated between hBN crystals~\cite{Dea10,Wan13} by a contamination-free van-der-Waals dry transfer process~\cite{Ban15}. Prior to transfer, the graphene/copper foil is stored under ambient conditions for a few days to permit the oxidation of the copper along the copper-to-graphene interface. This step is of utmost importance, as it weakens the adhesion between graphene and copper. Subsequently, the graphene is purely mechanically lifted from the copper foil with the help of an exfoliated hBN crystal placed on a stack of polymethyl methacrylate (PMMA) and polyvinyl alcohol (PVA), that is supported by a stamp of polydimethylsiloxane (PDMS). As illustrated in Figure 1b, the graphene gets detached along the edges of the hBN crystal when being delaminated. Thereafter, the graphene/hBN stack is placed on a second hBN crystal, which was exfoliated on a Si$^{++}$/SiO$_2$ substrate beforehand (see illustration of hBN/graphene/hBN sandwich in Figure 1c). Finally, the PDMS is mechanically removed and the remaining PVA and PMMA are dissolved in water and acetone, respectively. A typical hBN/graphene/hBN heterostructure is shown in Figure 1d. As the graphene is lifted from the copper foil using exfoliated hBN, the sample size is limited ultimately by the size of the exfoliated hBN crystals. In order to significantly increase the achievable device size, we exfoliate hBN using two stamps of PDMS. This process yields clean hBN crystals that are substantially larger than those exfoliated with scotch tape.

After the assembly of the sandwich, we perform spatially-resolved confocal Raman spectroscopy measurements over the entire sandwich structure~\cite{Gra07}, to prove the structural quality of the specimen. A typical Raman spectrum, recorded with a laser excitation at a wavelength of 532~nm, is shown in Figure~1f. The spectrum exhibits the characteristic hBN peak as well as the G and 2D peaks of graphene. The absence of a D-peak around $\omega_{\mathrm{D}}$~=~1345~$\mathrm{cm}^{-1}$ indicates a very low density of lattice defects in the graphene. Moreover, the 2D-peak exhibits a very small full-width-at-half-maximum (FWHM) of $\Gamma_{\mathrm{2D}}=16.5~\mathrm{cm}^{-1}$. This value is comparable with those reported for high mobility samples obtained from both exfoliated and CVD graphene~\cite{Neu14, Fer06, For13, Gra07, Ban15}. The positions of the G- and 2D-peaks around $\omega_{\mathrm{G}}$~=~$1,582~\mathrm{cm}^{-1}$ and $\omega_{\mathrm{2D}}$~=~$2,686~\mathrm{cm}^{-1}$, respectively, point to an overall low doping concentration of graphene.~\cite{Lee12} Figure~1e shows the spatial map of the width of the 2D peak, $\Gamma_{\mathrm{2D}}$, of the hBN/graphene/hBN sample presented in Figure~1d. In agreement with previous studies,\cite{Neu14, Ban15} $\Gamma_{\mathrm{2D}}$ shows the lowest values in regions where the graphene is fully encapsulated between hBN from both the top and bottom surface (see region I inside the dotted line in Figures~1d and 1e). In contrast, $\Gamma_{\mathrm{2D}}$ is slightly larger in areas where graphene is protected by hBN on just one surface (region II in Figure~1d). The histogram of the values of $\Gamma_{\mathrm{2D}}$ recorded in region I, (the encapsulated area) is shown in the inset of Figure~1f. It shows exceptionally low $\Gamma_{\mathrm{2D}}$ values, thus indicating that the sample is characterized by very small nanometer-scale strain variations over the entire graphene layer~\cite{Neu14}. Moreover, low values of $\Gamma_{\mathrm{2D}}$ are a hallmark for putatively high carrier mobilities in a charge transport device, which are needed for ballistic transport over long distances.~\cite{Cuo14, Ban15} To verify this, we fabricated devices from these CVD graphene heterostructures. The sandwiches are patterned by electron beam lithography and reactive ion etching using SF$_6$/Ar plasma into Hall cross structures or squares. We followed the scheme by Wang et al. to fabricate one dimensional Cr/Au (5~nm/95~nm) side contacts.~\cite{Wan13} In the following, we study a cross-shaped structure with a bar width of 1~$\mu$m and large square-shaped devices with an edge length of 20~$\mu$m.



\begin{figure*}[t!]
\centering
\includegraphics[draft=false,keepaspectratio=true,clip,%
                   width=0.98\linewidth]%
                   {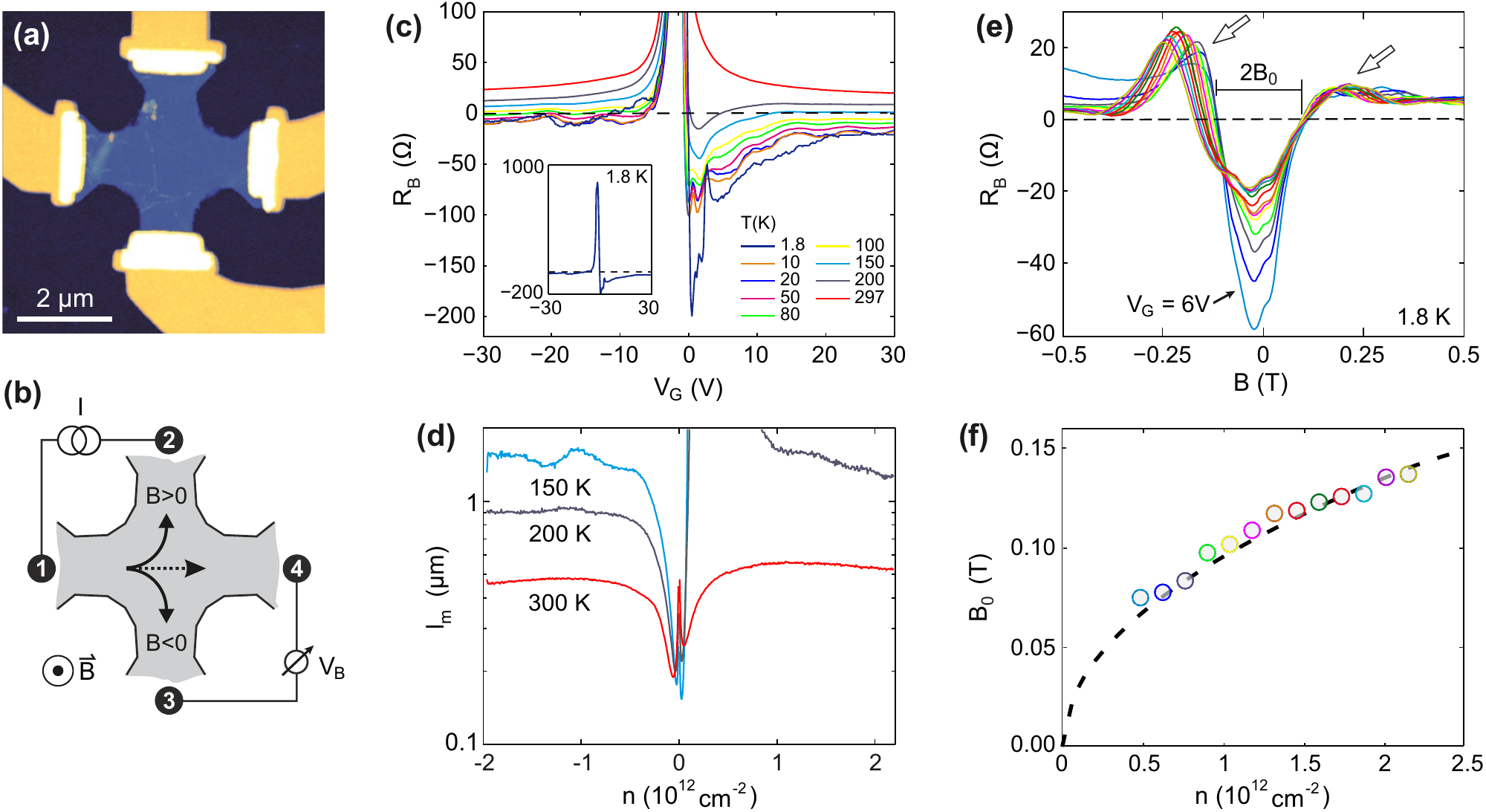}
\caption[fig02]{\textbf{(a)} AFM image of a hBN/CVD-graphene/hBN Hall cross device with Cr/Au side contacts. \textbf{(b)} Illustration of electrical wiring. The current is driven from contact 1 to 2, while the bend voltage $V_B$ is measured between contacts 3 and 4. An out-of plane magnetic field $B$ causes cyclotron motion of charges into opposite directions when reversing the magnetic field polarity. \textbf{(c)} Bend resistance $R_{\mathrm{B}}$ measured at various temperatures and zero magnetic field. Ballistic transport from contact 1 to contact 4 is seen for $R_{\mathrm{B}} < 0$. Inset: complete back gate characteristic at 1.8~K. \textbf{(d)} Mean free path calculated from the diffusive regime as function of the charge carrier concentration for 150, 200 and 300~K. \textbf{(e)} Magnetic field dependence of the bend resistance for different back gate voltages [6 to 30~V in 2~V steps; cf. colored data points in panel (f)]. Depending on the magnetic field polarity the charge carriers are deflected into contacts 2 or 3, resulting in positive $R_{\mathrm{B}} > 0$ values above a threshold field $B_0$. \textbf{(f)} $B_0$ as function of charge carrier density. The black dashed line is the expected dependence for a cyclotron radius of 1~$\mu$m (see main text). The colors of the circles correspond to the colors of the respective traces in panel (e).
}
\label{fig02}
\end{figure*}

In Figure~2, we show an atomic force microscopy (AFM) image of a contacted cross-shaped device. The hBN/graphene/hBN stack is seen from the top as the blue inner region. A straightforward measurement scheme for probing ballistic transport is to determine the so-called bend resistance $R_{\mathrm{B}}$ (see illustration in Figure~2b). In this scheme, we apply a current $I$ between the two neighboring contacts 1 and 2 and measure the voltage drop $V_{\mathrm{B}}$ over the two remaining contacts 3 and 4 with the bend resistance $R_{\mathrm{B}}~=~V_{\mathrm{B}}/I$~\cite{Hir91,May11}. A negative bend resistance indicates a ballistic over-shoot of charge carriers from the injection electrode (1) into the detection electrode (4). Figure~2c shows $R_{\mathrm{B}}$ of the device in Figure~2a as function of back gate voltage $V_{\mathrm{G}}$ (which is applied to the
underlying Si$^{++}$/SiO$_2$ substrate) at zero magnetic field (B = 0), and  in a temperature range between 1.8 and 300~K. At low temperatures, the bend resistance becomes negative for both electron and hole doping, with a distinct minimum in the electron regime close to the charge neutrality point (see Figure~2c). This negative bend resistance unambiguously demonstrates ballistic transport of charge carriers from the source contact 1 to the opposite voltage contact 4 without being scattering (see dotted line in Figure~2b). We conclude that the charge carriers exhibit an elastic mean free path $l_{\mathrm{m}}$ exceeding $1~\mu$m. Such a behavior has previously been observed in similar devices fabricated from exfoliated graphene.\cite{May11} The bend resistance exhibits a strong temperature dependence and becomes positive for almost all charge carrier densities at 200~K (see gray curve in Figure~2c) indicating the transition to the regime of diffusive charge transport, governed by electron-phonon scattering. This limits the room temperature mobility of graphene to values around 130,000~cm$^2$/(Vs) at carrier densities around 10$^{12}$~cm$^{-2}$.\cite{Che08,Hwa08} At all temperatures diffusive transport prevails at very low carrier densities near the charge neutrality point at $n=0$, as the mean free path depends on the charge carrier density, $l_m\propto \sqrt{n}$. The charge neutrality point is identified by the position of the maximum bend resistance at $V_{\mathrm{G,0}}=-1.2$~V at 1.8~K (see inset of Figure~2c).

Charge carrier mobilities can be estimated using the relation $\mu~=~2e\sqrt{\pi}l_{\mathrm{m}}/(h\sqrt{n_{\mathrm{th}}})$, where $n_{\mathrm{th}}$ is the threshold carrier density in which the system enters the ballistic transport regime, which is reached when $R_{\mathrm{B}}$ becomes negative.\cite{May11,Wan13} At 1.8~K, we estimate $\left|n_{\mathrm{th}}\right|~=~3.3\times 10^{11}$~$\mathrm{cm}^{-2}$ for holes at a back gate voltages $V_{\mathrm{G,th}}$~=$-6.1$~V using $n_{\mathrm{th}}$~=~$\alpha$($V_{\mathrm{G,th}}$-$V_{\mathrm{G,0}}$), where $\alpha$~=~$6.9\times 10^{10}$~cm$^{-2}$V$^{-1}$ is the back gate lever arm. As discussed above, bend resistance measurements indicate that at this temperature the mean free-path is at least of the order of the sample size. Taking $l_{\mathrm{m}}=1~\mu$m, we obtain a charge carrier mobility of $\mu$~=~$147,000~\mathrm{cm}^2$/(Vs). The estimated carrier mobility is comparable to values recently reported for high mobility CVD graphene encapsulated in hBN~\cite{Ban15}. In addition, we calculate the density-dependent charge carrier mean free path from the diffusive part of the back gate traces at 150, 200 and 300~K (Figure~2d). To do so, we apply the Drude formula $\sigma=en\mu$. In order to obtain the conductivity $\sigma$ we use the van-der-Pauw formula and the high symmetry of the diffusive part of the back gate traces for different measurement configurations, following Wang et al.\cite{Wan13}. The mean free path can then be calculated using $l_m=(h/2e)\mu\sqrt{n/\pi}$. Also this analysis gives a mean free-path, comparable with the system size, i.e. $l_m$=1~$\mu$m, which is in agreement with the values extracted from the appearance of a negative bend resistance.

\begin{figure*}[t!]\centering
\includegraphics[draft=false,keepaspectratio=true,clip,%
                   width=0.95\linewidth]%
                   {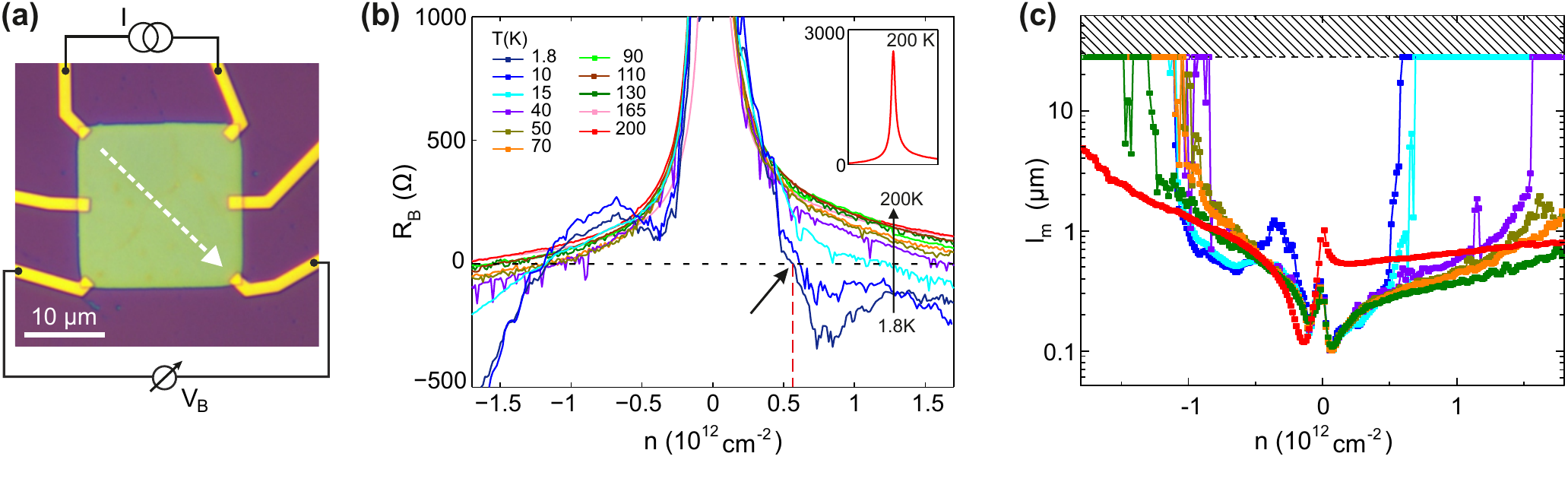}
\caption[fig03]{
 \textbf{(a)} Optical image of a square-shaped hBN/CVD-graphene/hBN device with Cr/Au side contacts. Ballistic transport is probed along the diagonal (dashed line) over 28~$\mu$m. \textbf{(b)} Bend resistance as function of charge carrier concentration for temperatures ranging from 1.8 to 200~K. Inset: complete $R_{\mathrm{B}}$ trace at 200~K. Charge transport is fully diffusive at this temperature, as seen by the positive $R_{\mathrm{B}}$. \textbf{(c)} Elastic mean free path calculated from selected $R_{\mathrm{B}}$ traces.
}
\label{fig03}
\end{figure*}

To further explore the ballistic nature of charge transport, we next apply a perpendicular magnetic field $B$ to the device and plot $R_{\mathrm{B}}$ vs.~$B$ in Figure 2e, for different back gate voltages ranging from $V_{\mathrm{G}}$=6~V to $V_{\mathrm{G}}=$~30~V. For all traces, $R_{\mathrm{B}}$ changes sign at moderate magnetic field strengths for both magnetic field polarities. This behavior is expected, in the ballistic regime, the electrons get deflected and travel on cyclotron orbits. If the cyclotron radius  $r_c~=~\hbar\sqrt{\pi n}/(eB)$ becomes small enough, the electrons can no longer ballistically reach probe contact 4, thus resulting in a positive $R_B$. The increased bend resistances close to $\pm$250~mT can be well attributed to the deflection of charge carriers in the other probing electrode (contact 3) and the drain electrode (contact 2) for both negative and positive magnetic fields, respectively (see right and left arrows in Figure~2e). To quantify the deflection of the carriers due to the magnetic field, we define the \emph{minimum deflection field} $B_{\mathrm{0}}$ as the average magnetic field resulting in zero bend resistance for both magnetic field polarities (see Figure~2e). In Figure~2f we show $B_{\mathrm{0}}$ as a function of electron density $n$ for all traces shown in Figure~2e. The dashed line, which represents the expected dependence for a cyclotron radius of 1~$\mu$m - consistently with the geometry of the sample -, matches very well to the experimental values.


As the Hall cross device shows ballistic transport even at high temperatures, we expect to observe ballistic transport in CVD-grown graphene devices with lateral dimensions much larger than 1~$\mu$m. To prove this, we fabricated large square-shaped devices of an edge length $\ell$=20~$\mu$m  with electrical side contacts at their corners and edges (see optical image in Figure~3a). For ballistic transport studies we use the wiring scheme in Figure~3a, as well as a configuration which is rotated by 90$^{\circ}$ with respect to the one shown. The current $I$ is injected through two neighboring contacts at the upper corners, while the voltage $V_{\mathrm{B}}$ is measured at the two opposing lower corners of the device. Figure~3b shows the corresponding bend resistance $R_{\mathrm{B}}~=~V_{\mathrm{B}}/I$ as function of charge carrier density $n$ for temperatures ranging from 1.8~K to 200~K. Ballistic transport is again identified by the negative bend resistance for electron and hole transport regimes, which occurs for both: the measurement configuration shown in Figure~3a as well as the wiring configuration that is rotated by 90$^{\circ}$ (data not shown). Similar to the discussion above, it can be explained by charge carriers traveling ballistically along the diagonal of the device (see dashed white line in Figure~3a), resulting in a negative voltage drop at low temperatures. The threshold carrier density for electrons is  $n_{\mathrm{th}}=6\times 10^{11}~\mathrm{cm}^{-2}$ and $n_{\mathrm{th}}=-1.3\times 10^{12}~\mathrm{cm}^{-2}$ for holes at $T=1.8$~K, respectively. From the lowest threshold carrier density for ballistic transport on the electron side $n_{\mathrm{th}}=6\times 10^{11}~\mathrm{cm}^{-2}$ (see arrow in Figure 3b) and a mean free path of 28 $\mu$m we estimate electron mobilities exceeding $3\times~10^6$ cm$^2$/(Vs) at T = 1.8 K.

For electrons, transport becomes diffusive ($R_{\mathrm{B}}>0$) around 40~K (dark purple curve in Figure~3b), while it remains ballistic even for temperatures up to 130~K for the largest hole concentrations. Since it is well known that the mean free-path increases at lower temperatures, a negative bend resistance at 130~K is a clear indication that at lower temperatures the mean free-path must well exceed the device diagonal $\ell\sqrt{2}$~=~28~$\mu$m.

Similar conclusions can be drawn from Figure~3c, where we plot the mean free path calculated from the van-der-Pauw conductivity through $l_{\mathrm{m}}=\sigma h/(2 e^2 k_{\mathrm{F}})$, with $k_{\mathrm{F}}=\sqrt{\pi n}$ in the diffusive transport regime, as function of $n$ for selected measurements presented in Figure~3b.
As expected, there is an overall strong increases of $l_{\mathrm{m}}$ away from the charge neutrality point at all temperatures. Whenever $l_{\mathrm{m}}$ reaches the sample diagonal of 28~$\mu$m (see dashed line in Figure~3c) transport becomes fully ballistic. Although we cannot extract $l_{\mathrm{m}}$ in this regime, our data suggest that the mean free path in our CVD-graphene exceeds by far 28~$\mu$m. Determining the actual value of $l_m$ at low temperature and high carrier density would require samples of much larger dimensions than what is achievable with the present technology, which is limited by the size of the exfoliated hBN crystals used to encapsulate graphene. In fact, even though, we improved on hBN exfoliation using PDMS stamps, resulting in hBN flakes of several tens of microns - these remain more than one order of magnitude smaller than the graphene flakes, which can be easily grown on the scale of one millimeter with high structural quality.
A possible way to overcome this shortcoming would be the use of high quality CVD grown crystals of hBN, which can also have sizes of several hundred micrometers~\cite{Can15} and in future may also be available with different thicknesses.

In summary, we demonstrate that the electronic quality of CVD-grown graphene supports
ballistic transport over distances larger than 28 $\mu$m at low temperature, with a mean free-path
that remains of the order of  1 $\mu$m even up to 200~K. A mean free path of 28~$\mu$m is, to the best of our knowledge, the largest value so far achieved in graphene in general\cite{Wan13} and almost two orders of magnitude larger than what has been demonstrated in CVD graphene by Calado et al. so far\cite{Cal14}. Similar to the results of Wang et al.\cite{Wan13}, who observed a mean free path of around 20~$\mu$m in exfoliated graphene encapsulated in hBN, we are currently limited in our mean free path by the device size. This unambiguously shows that even for our relatively large devices, there is no intrinsic quality difference between exfoliated and synthetic graphene, as long as a transfer process induced quality decrease is avoided.
These results represent a fundamental step towards Dirac-fermion optic applications such as Veselago lenses, high-frequency Klein-tunneling transistors or ballistic electron
waveguides~\cite{Ric15}. All these applications require long mean free paths in large samples surrounded by ultra-thin oxides as a prerequisite. With our work we show that this conditions can be met by current CVD-based graphene fabrication technologies~\cite{Ban15}, at least for samples with sizes of the order of hundreds of $\mu$m. Even larger samples will require significant improvement in the fabrication of large and thin hBN flakes, however, progresses in this direction are already underway~\cite{Can15}.




The authors thank F. Haupt and R. McNeil for comments on the manuscript and S. Kuhlen for help on the figures. Support by the Helmholtz-Nanoelectronic-Facility (HNF), the DFG (SPP-1459), the ERC (GA-Nr. 280140), and the EU project Graphene Flagship (contract no. NECT-ICT-604391) are gratefully acknowledged.\\

\end{document}